\documentclass[floatfix,aps,showpacs,twocolumn,nofootinbib]{revtex4}
\usepackage{graphicx}

\begin{document}

\title{Monte Carlo reconstruction of the inflationary potential}
\author{Richard Easther} \email{easther@physics.columbia.edu}
\author{William H.\ Kinney} \email{kinney@physics.columbia.edu} \affiliation{
Institute for Strings, Cosmology and Astroparticle Physics\\
Columbia University\\
550 W. 120th St., New York, NY 10027
}
\date{October 15, 2002}

\begin{abstract}
We present Monte Carlo reconstruction, a new method for ``inverting''
observational data to constrain the form of the scalar field potential
responsible for inflation. This stochastic technique is based on the
flow equation formalism and has distinct advantages over
reconstruction methods based on a Taylor expansion of the potential.
The primary ansatz required for Monte Carlo reconstruction is simply
that inflation is driven by a single scalar field.  We also require a
very mild slow roll constraint, which can be made arbitrarily weak
since Monte Carlo reconstruction is implemented at arbitrary order in
the slow roll expansion.  While our method cannot evade fundamental
limits on the accuracy of reconstruction, it can be simply and
consistently applied to poor data sets, and it takes advantage of the
attractor properties of single-field inflation models to constrain the
potential outside the small region directly probed by observations. We
show examples of Monte Carlo reconstruction for data sets similar to
that expected from the Planck satellite, and for a hypothetical
measurement with a factor of five better parameter discrimination than
Planck.
\end{abstract}

\maketitle

\section{Introduction}

Given a detailed model of physics at the Grand Unified Theory (GUT)
scale that supports inflation one can deduce many of the overall
properties of the universe from first principles.  However, new
strides in precision cosmology are likely to present us with the
inverse problem: given an accurate measurement of the primordial
perturbation spectrum, can we determine the particle physics that
drove inflation?  This question leads to ``reconstruction'' --
rebuilding a portion of the inflationary potential from observations
-- whose apotheosis is the work of Lidsey et al. \cite{lidsey95}. The
conclusion of this effort was that while a limited reconstruction may
be possible, it is overly optimistic to hope to deduce more than the
rough form of a small piece of the inflaton potential on the basis of
astrophysical observations alone.

Despite this apparent setback, the prospect of being able to probe
ultra-high energy particle physics via astrophysical measurements
remains tempting. In this paper we present a new approach to the task,
{\it Monte Carlo reconstruction\/}.  Unlike previous workers, we
tackle the problem stochastically, by generating large numbers of
inflationary models and their associated potentials and identifying
those that lead to a spectrum of primordial perturbations which fits
inside a specified window of parameter space.  We sacrifice the
prospect of unambiguously reconstructing the inflationary potential,
but instead map out the ensemble of viable potentials. We can then
assess the extent to which these potentials resemble one another, and
thus the degree of ambiguity in the ``reconstruction''.

The starting point for Monte Carlo reconstruction is the Hubble slow
roll expansion \cite{liddle94}, which leads to a set of first order
``flow'' equations \cite{hoffman00,kinney02} that describes the
inflationary dynamics. The Hubble slow roll expansion can be
systematically continued to arbitrary order but, as we explain below,
the truncated expansion actually yields an exact solution of the full
equations, drawn from a subset of the overall
solution space. We select a ``universe'' by assigning random values to
each term in the truncated slow roll expansion, and integrate the flow
equations forward to the end of inflation (or far enough into the
future to show that the solution approaches an inflationary
attractor). We can then determine the epoch that corresponds to
cosmological structure formation and thus the predicted parameters of
the primordial spectrum, allowing us to select the models which fit
into a pre-determined window of parameter space.

As with any other reconstruction program, we face limitations dictated
by the physics of inflation itself. Since the field rolls slowly, it
traverses only a small portion of the potential during the epoch in
which the primordial perturbations are laid down.  Consequently, even
if all the potentials we recover for a specific set of cosmological
parameters overlap in the region where the cosmologically relevant
portion of the spectrum is laid down, they will still diverge
outside this region.  However, because we reconstruct the entire
potential, we can estimate the degree to which the reconstructed
potentials diverge from one another.

In the following section we summarize the inflationary dynamics and
the flow equations, and in Section 3 apply these to spell out the
process of Monte Carlo reconstruction in detail.  We present the
results of a variety of simulations in Section 4, and test how
accurately Monte Carlo reconstruction can recover a known inflationary
potential. As expected, a fully consistent reconstruction requires the
unambiguous detection of tensor modes in the CMB, but even in the
absence of a detectable tensor signal we can still recover some
information about the inflationary potential.  If we assume
the error bars on the scalar spectral index and tensor-to-scalar ratio
that are expected for the Planck mission, the underlying potential
cannot be unambiguously recovered. However, if we decrease these error
bars (perhaps by adding information from large scale structure
surveys, or better measurements of CMB polarization) by a factor of 5,
then we can start to recover a $\phi^n$ potential. Our
conclusions are summarized in Section 5.

\section{Inflation and flow}
\label{secflowreview}

We consider inflation driven by a single homogeneous scalar field
$\phi$ (the {\em inflaton}) with potential $V(\phi)$ and
equation of motion
\begin{equation}
\ddot \phi + 3 H \dot \phi + V'(\phi) = 0,\label{eqequationofmotion}
\end{equation}
where $H \equiv (\dot a / a)$ is the Hubble parameter. We assume
a spatially flat, homogeneous and isotropic universe where the inflaton
field is the only contribution to the energy-momentum tensor, so that the
Einstein field equations have the familiar form
\begin{equation}
H^2 = \left({\dot a \over a}\right)^2 = {8 \pi \over 3 m_{\rm Pl}^2}
 \left[V(\phi) + {1 \over 2}
 \dot\phi^2\right],\label{eqbackgroundequation1}
\end{equation}
and
\begin{equation}
{\ddot a \over a} = {8  \pi \over 3 m_{\rm Pl}^2}
 \left[V(\phi) - \dot\phi^2\right].\label{eqbackgroundequation2}
\end{equation}
Here $m_{\rm Pl} = G^{-1/2} \simeq 10^{19}\,{\rm GeV}$ is the Planck
mass. These background equations, along with the equation of motion
(\ref{eqequationofmotion}), form a coupled set of differential
equations that describe the evolution of the universe. The limit
$\dot\phi = 0$ corresponds to a de Sitter universe, with the scale
factor increasing exponentially in time
\begin{eqnarray} &&H = \sqrt{\left({8 \pi \over 3 m_{\rm Pl}^2}\right)
V(\phi)} = {\rm const},\cr &&a \propto e^{H t}.
\end{eqnarray}
In general $H$ is not exactly constant, but varies as the field $\phi$
evolves along the potential $V(\phi)$. A powerful way of
describing the dynamics of an inflationary universe with a varying
field (and non-trivial potential) is to express the Hubble parameter
as a function of the field $\phi$, $H = H(\phi)$, which is
consistent provided $\phi$ is monotonic in time. The equations of
motion become \cite{grishchuk88,muslimov90,salopek90,lidsey95}
\begin{eqnarray} \dot\phi
&=& -{m_{\rm Pl}^2 \over 4 \pi} H'(\phi),\cr
\left[H'(\phi)\right]^2 - {12 \pi \over m_{\rm Pl}^2}
H^2(\phi) &=& - {32 \pi^2 \over m_{\rm Pl}^4}
V(\phi).\label{eqbasichjequations}
\end{eqnarray}
These are completely equivalent to the second-order equation of motion
(\ref{eqequationofmotion}). The second of the above equations is
referred to as the {\it Hamilton-Jacobi} equation, and can be written
in the useful form
\begin{equation} 
H^2(\phi) \left[1 - {1\over 3}
\epsilon(\phi)\right] =  \left({8 \pi \over 3 m_{\rm Pl}^2}\right)
 V(\phi),\label{eqhubblehamiltonjacobi}
\end{equation}
where $\epsilon$ is defined to be
\begin{equation}
\epsilon \equiv {m_{\rm Pl}^2 \over 4 \pi} \left({H'(\phi) \over
 H(\phi)}\right)^2.\label{eqdefofepsilon}
\end{equation}
The physical meaning of $\epsilon$ can be seen by expressing Eq.
(\ref{eqbackgroundequation2}) as
\begin{equation}
\left({\ddot a \over a}\right) = H^2 (\phi) \left[1 -
 \epsilon(\phi)\right],
\end{equation}
so that the condition for inflation $(\ddot a / a) > 0$ is given by
$\epsilon < 1$. The scale factor is given by
\begin{equation}
a \propto e^{N} = \exp\left[\int_{t_0}^{t}{H\,dt}\right],
\end{equation}
where the number of e-folds $N$ is
\begin{equation}
N \equiv \int_{t}^{t_e}{H\,dt} = \int_{\phi}^{\phi_e}{{H \over
\dot\phi}\,d\phi} = {2 \sqrt{\pi} \over m_{\rm Pl}}
\int_{\phi_e}^{\phi}{d\phi \over
\sqrt{\epsilon(\phi)}}.\label{eqdefofN}
\end{equation}
It is convenient to use $N$ as the measure of time during
inflation. We take $t_e$ and $\phi_e$ to be the time and field value
at end of inflation. Therefore $N$ is defined as the number of e-folds
before the end of inflation and increases as one goes {\em backward}
in time, $d t > 0 \Rightarrow d N < 0$. The sign convention for
$\sqrt{\epsilon}$ must be applied carefully, and we take it to have
the same sign as $H'(\phi)$:
\begin{equation}
\sqrt{\epsilon} \equiv + {m_{\rm PL} \over 2 \sqrt{\pi}} {H' \over H}.
\end{equation}
Liddle, Parsons and Barrow use this as the starting point for an
infinite hierarchy of slow roll parameters \cite{liddle94}:
\begin{eqnarray}
\sigma &\equiv& {m_{\rm Pl} \over \pi} \left[{1 \over 2} \left({H'' \over
 H}\right) -
\left({H' \over H}\right)^2\right],\cr
{}^\ell\lambda_{\rm H} &\equiv& \left({m_{\rm Pl}^2 \over 4 \pi}\right)^\ell
{\left(H'\right)^{\ell-1} \over H^\ell} {d^{(\ell+1)} H \over d\phi^{(\ell +
1)}}.
\end{eqnarray}
The evolution of these parameters during inflation is
determined by a set of ``flow'' equations \cite{hoffman00,kinney02},
\begin{eqnarray}
{d \epsilon \over d N} &=& \epsilon \left(\sigma + 2
\epsilon\right),\cr {d \sigma \over d N} &=& - 5 \epsilon \sigma - 12
\epsilon^2 + 2 \left({}^2\lambda_{\rm H}\right),\cr {d
\left({}^\ell\lambda_{\rm H}\right) \over d N} &=& \left[
\frac{\ell - 1}{2} \sigma + \left(\ell - 2\right) \epsilon\right]
\left({}^\ell\lambda_{\rm H}\right) + {}^{\ell+1}\lambda_{\rm
H}.\label{eqfullflowequations}
\end{eqnarray}
Here the time variable is the number of e-folds $N$, where
\begin{equation}
{d \over d N} = {d \over d\log a} = { m_{\rm Pl} \over 2 \sqrt{\pi}}
\sqrt{\epsilon} {d \over d\phi}.
\end{equation}
The derivative of a slow roll parameter at a given order is higher
order in slow roll. Taken to infinite order, this set of equations
completely specifies the cosmological evolution, up to the
normalization of the Hubble parameter $H$.  When truncated at {\em
finite\/} order, by assuming that the ${}^\ell\lambda_{\rm H}$ are all
zero above some fixed value of $\ell$, the solution of the flow
equations still yields an {\em exact\/} solution of the background
equations, albeit one that is drawn from a special subset of the
overall solution space.

\section{Monte Carlo reconstruction}

The inflationary dynamics are fully specified by the values of the
slow roll parameters $[\epsilon,\sigma,{}^\ell\lambda_{\rm H}]$ at a
fixed time, which serve as initial conditions for the flow equations,
truncated at order $M$ in the slow roll expansion. We now show that
this information determines the inflationary potential $V(\phi)$, up
to a constant multiplier.
The starting point is the Hamilton-Jacobi equation,
\begin{equation}
V(\phi) = \left({3 m_{\rm Pl}^2} \over 8 \pi\right)
H^2(\phi) \left[1 - {1\over 3}
\epsilon(\phi)\right].\label{eqHJpotential}
\end{equation}
We have $\epsilon(N)$ trivially from the flow equations. In
order to calculate the potential, we need to determine
$H(N)$ and $\phi(N)$. With $\epsilon$ known,
$H(N)$ can be determined by inverting the definition of
$\epsilon$, Eq.  (\ref{eqdefofepsilon}):
\begin{equation}
{1 \over H} {d H \over d N} = \epsilon.
\end{equation}
This can be viewed as the lowest-order member of the system of flow
equations (\ref{eqfullflowequations}). Similarly, $\phi(N)$ follows
from the first Hamilton-Jacobi equation (\ref{eqbasichjequations}):
\begin{equation}
{d \phi \over d N} = {m_{\rm PL} \over 2 \sqrt{\pi}}
\sqrt{\epsilon}.
\end{equation}
Using these equations and Eq.~(\ref{eqHJpotential}), the form of the
potential can then be fully reconstructed from the numerical solution
for $\epsilon(N)$.\footnote{Expressing the potential in this way is certainly
not new. This is the basis of ``functional reconstruction'' method of 
Refs. \cite{hodges90,copeland93}, and
well as the ``Stewart-Lyth inverse problem'' of Ref.
\cite{beato00}.}
The only necessary observational input is the
normalization of the Hubble parameter $H$, which enters the above
equations as an integration constant.  Here we use the simple
condition that the density fluctuation amplitude (as determined by a
first-order slow roll expression) be of order $10^{-5}$,
\begin{equation}
{\delta \rho \over \rho} \simeq {H \over 2 \pi m_{\rm Pl} \sqrt{\epsilon}} =
 10^{-5}.
\end{equation}
A more sophisticated treatment would perform a full normalization to
the COBE CMB data \cite{bunn94,stompor95}.  The value of the field,
$\phi$, also contains an arbitrary, additive constant.

Given a solution to the flow equations it is straightforward to
determine the observational predictions for that model, even before
the potential is computed.  For instance, measurements of CMB
anisotropies \cite{dodelson97,kinney98} may determine the
tensor/scalar ratio $r$, the spectral index $n$, and the ``running''
of the spectral index $d n / d\log k$, and we focus on these
quantities here.  To lowest order, the relationship between the slow
roll parameters and the observables is especially simple: $r =
\epsilon$, $n - 1 = \sigma$, and $d n / d \log k = 0$. To second order
in slow roll, the observables are given by
\cite{liddle94,stewart93},
\begin{equation}
r = \epsilon \left[1 - C \left(\sigma + 2
 \epsilon\right)\right],\label{eqrsecondorder}
\end{equation}
for the tensor/scalar ratio, and 
\begin{equation}
n - 1 = \sigma - \left(5 - 3 C\right) \epsilon^2 - {1 \over 4} \left(3
- 5 C\right) \sigma \epsilon + {1 \over 2}\left(3 - C\right)
\left({}^2\lambda_{\rm H}\right)\label{eqnsecondorder}
\end{equation}
for the spectral index. The constant  $C \equiv 4 (\ln{2} +
\gamma)$, where $\gamma \simeq 0.577$ is Euler's
constant. Derivatives with respect to wavenumber $k$ can be expressed
in terms of derivatives with respect to $N$ as \cite{liddle95}
\begin{equation}
{d \over d N} = - \left(1 - \epsilon\right) {d \over d \ln k},
\end{equation}
The scale dependence of $n$ is then  given by the simple expression
\begin{equation}
{d n \over d \ln k} = - \left({1 \over 1 - \epsilon}\right) {d n \over d N},
\end{equation}
which can be evaluated to third order in slow roll by using
Eq.~(\ref{eqnsecondorder}) and the flow equations. The final result
following the evaluation of a particular path in $M$ dimensional
``slow roll space'' is a point in ``observable parameter space'',
i.e. $(r,n,dn/d\log k)$, corresponding to the observational prediction
for that particular model.  This process can be repeated for a large
number of models, and used to study the attractor behavior of the
inflationary dynamics. In fact, the models cluster
strongly in the observable parameter space~\cite{kinney02}.

We are now in a position to ask the key question: given an
observational constraint, what dynamics -- and underlying inflationary
potential -- are compatible with that constraint?  We proceed by
generating a large number of ``universes'' with random values (in a
sense that is made explicit below) for the slow roll parameters.
Given an allowed region in the observable parameter space, centered on
specified values of $r$,$n$, and $dn/d\log k$ with a width given by
the error bars on these quantities, we can construct an ensemble of
inflationary potentials consistent with with the specified ``window''
in parameter space.  
This Monte Carlo approach to reconstruction differs from previously
used techniques \cite{lidsey95,grivell00,escalante02} in that it generates
an ensemble of potentials consistent with a given constraint, rather than
attempting to produce a parameterized form for the potential and then 
constrain those parameters.
The advantage of Monte Carlo reconstruction is that by
evaluating the flow equations to high enough order we need only very
mild {\em a priori\/} assumptions about the form of the potential. In
addition, by probing the parameter space in a uniform way, we obtain
what is in some sense a ``fair sample'' of the potentials consistent
with a given observational bound. Even so, we do not have a metric on
the space of initial conditions, so it is not possible to derive
statistical inferences from the ensemble of potentials.

The condition for the end of inflation is that $\epsilon =
1$. Integrating the flow equations forward in time will yield two
possible outcomes. One possibility is that the condition $\epsilon = 1$ may
be satisfied for some finite value of $N$, which defines the end of
inflation. We identify this point as $N=0$ so that the primordial
fluctuations are actually generated when $N \sim 50$. Alternatively,
the solution can evolve toward an inflationary attractor with $r = 0$
and $n > 1$, in which case inflation never stops.\footnote{See Ref.
\cite{kinney02} for a  detailed discussion of the fixed-point structure 
of the slow roll space.}
In reality, inflation must stop at some point, presumably via some sort 
of phase transition, such as the ``hybrid'' inflation
mechanism \cite{linde91,linde94,copeland94}. Here we make the simplifying
assumption that the observables for such models are the values at the 
late-time attractor.

\begin{figure*}
\centerline{\includegraphics[width=4.5in]{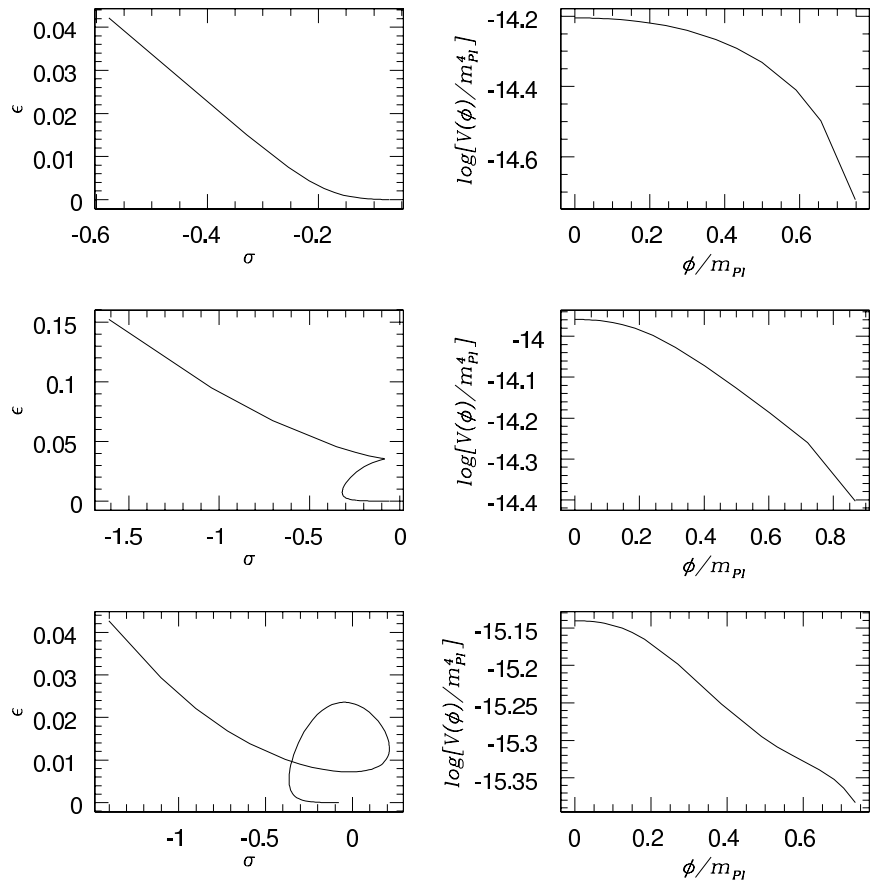}}  
\caption{Paths in the $(\sigma,\epsilon)$ plane (left column) and the
corresponding potentials $V(\phi)$ (right column) for $r \simeq 0.0$,
$n \simeq 0.93$, $d n / d \log k \simeq 0.0$.}
\end{figure*}

\begin{figure*}
\centerline{\includegraphics[width=4.5in]{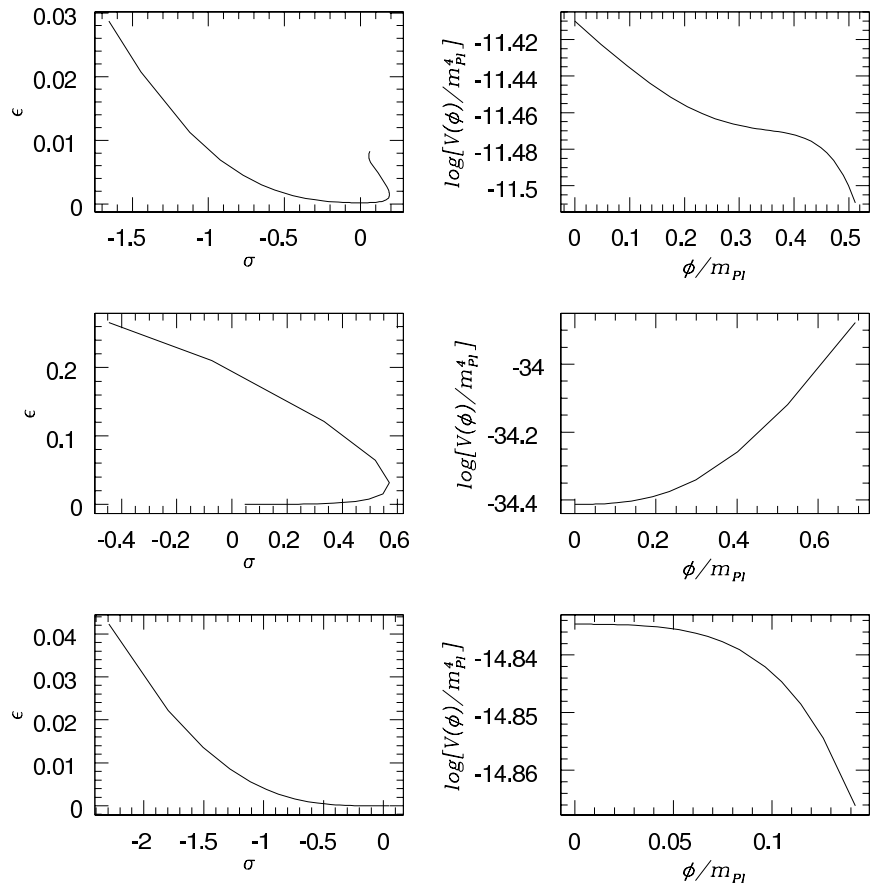}}  
\caption{Paths in the $(\sigma,\epsilon)$ plane (left column) and the
corresponding potentials $V(\phi)$ (right column) for a blue spectrum
$r \simeq 0.0$, $n \simeq 1.05$, $d n / d \log k \simeq 0.0$.} 
\end{figure*}

\begin{figure*}
\centerline{\includegraphics[width=4.5in]{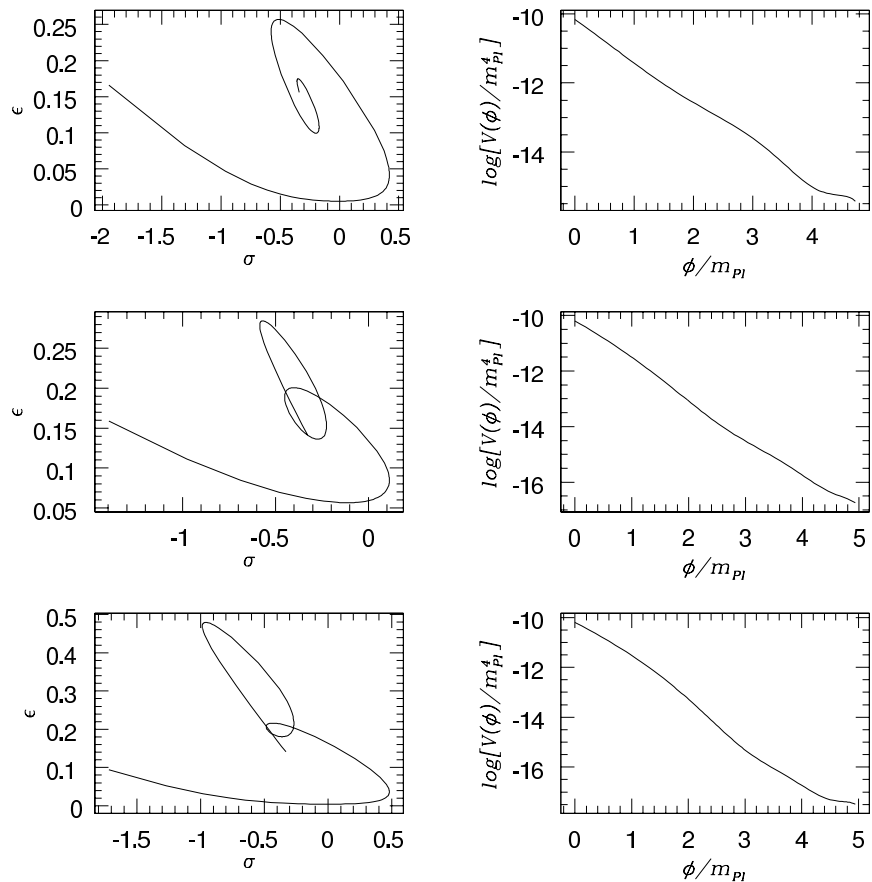}}  
\caption{Paths in the $(\sigma,\epsilon)$ plane (left column) and the
corresponding potentials $V(\phi)$ (right column) for $r \simeq 0.18$,
$n \simeq 0.6$, $d n d / \log k \simeq -0.02$. This parameter region
is observationally disfavored, but shows the complicated behavior
possible for solutions to the flow equations at higher order.} 
\end{figure*}

\begin{figure*}
\centerline{\includegraphics[width=4.5in]{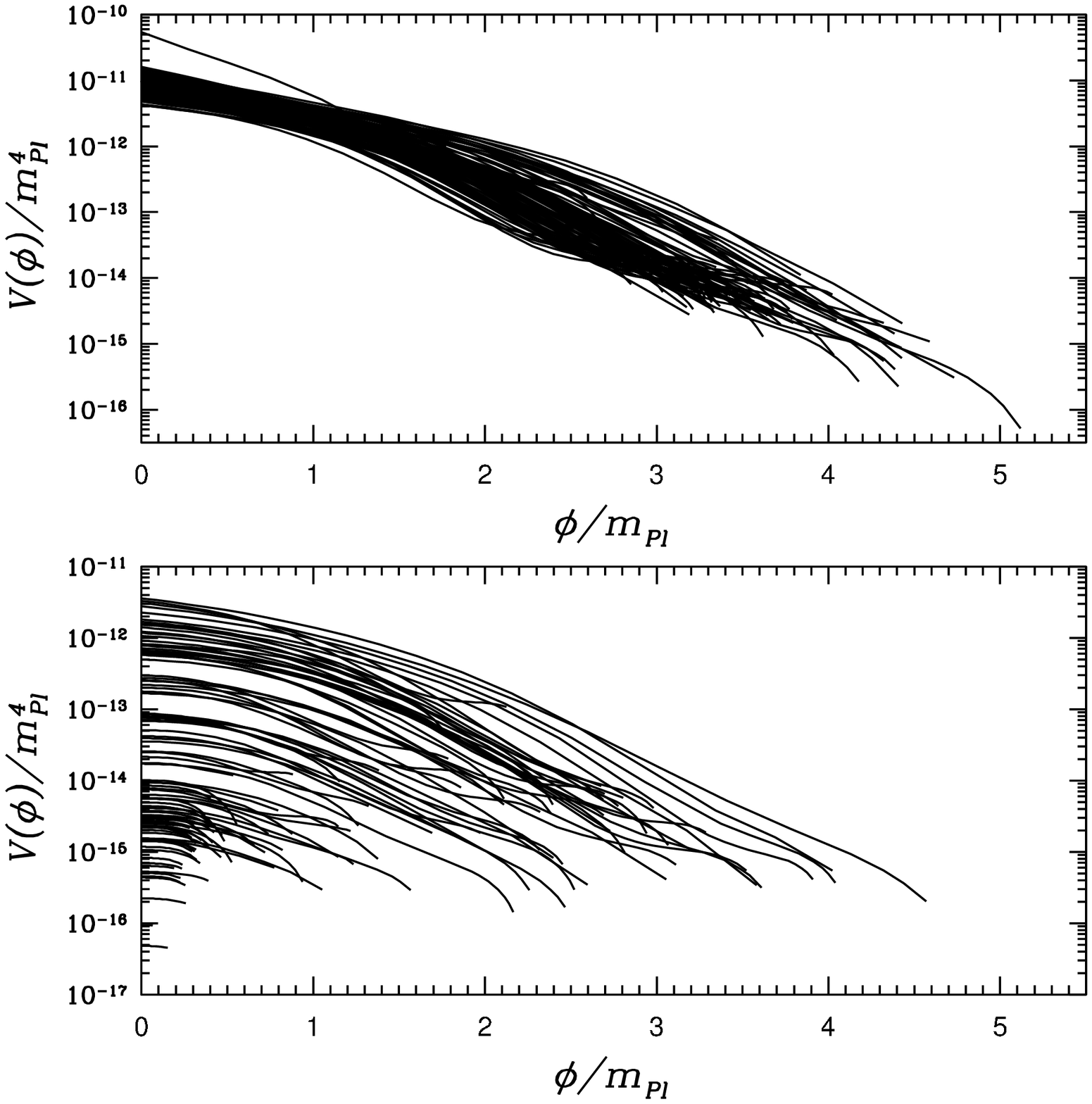}}  
\caption{The upper panel shows 100 reconstructed potentials, assuming
$r = 0.02 \pm 0.01$, $n = 0.95 \pm 0.01$, $d n / d \log{k} = 0.0 \pm
0.01$ (the errors bars expected from the Planck mission). This choice
implies that the tensor modes are unambiguously detected, and leads to
a tight constraint on the normalization of the potential. The field
$\phi$ is defined such that the observational parameters are
calculated at $\phi = 0$. The lower plot shows 100 reconstructed
potentials, for $r = 0.0 + 0.01$, $n = 0.93 \pm 0.01$, $d n / d \log k
= 0.0 \pm 0.01$. In this case, tensor modes are not resolved, and the
normalization of the potential is poorly constrained.  The one
anomalous potential in the top figure corresponds to a potential that,
by chance, has comparatively large slow roll parameters, but which
cancel in just the right way to produce the specified cosmological
spectrum.}
\end{figure*}

To summarize, the algorithm for Monte Carlo reconstruction is as follows:
\begin{enumerate}
\item Specify a ``window'' of parameter space: e.g. central values for
$n-1$, $r$ or $d n /d \ln{k}$ and their associated error bars.
\item Select a random point in slow roll space, 
$[\epsilon,\eta,{}^\ell\lambda_{\rm H}]$, truncated at order $M$ in
the slow roll expansion.
\item Evolve forward in time ($d N < 0$) until either (a) inflation ends
 ($\epsilon > 1$), or (b) the evolution reaches a late-time fixed
 point ($\epsilon = {}^\ell\lambda_{\rm H} = 0,\ \sigma = {\rm
 const.}$)
\item If the evolution reaches a late-time fixed point, calculate the
 observables $r$, $n - 1$, and $d n / d \ln k$ at this point.
\item If inflation ends, evaluate the flow equations backward $N$ e-folds from
 the end of inflation. Calculate the observable parameters at this
 point.
\item If the observable parameters lie within the specified window of
parameter  space, compute the potential and add this model to the ensemble 
of ``reconstructed'' potentials.
\item Repeat steps 2 through 6 until the desired number of models
have been found.
\end{enumerate}

In principle it is possible to carry out Monte Carlo reconstruction
with no assumptions about the convergence of the hierarchy of slow
roll parameters. In practice, the flow equations
(\ref{eqfullflowequations}) must be truncated at some finite order and
evaluated numerically. Moreover, for any given path in the parameter
space, we do not know {\it a priori} the correct number of e-folds $N$
at which to evaluate the observables, since this depends on details
such as the energy density during inflation and the reheat temperature
\cite{lidsey95}. Consequently, after truncating to order $M$ in slow
roll, we select our models' parameters randomly from the following
uniform distributions:
\begin{eqnarray}
N &=& [40,70]\cr
\epsilon &=& \left[0,0.8\right]\cr
\sigma &=& \left[-0.5,0.5\right]\cr
{}^2\lambda_{\rm H} &=& \left[-0.05,0.05\right]\cr
{}^3\lambda_{\rm H} &=& \left[-0.025,0.025\right],\cr
&\cdots&\cr
{}^{M+1}\lambda_{\rm H} &=& 0.\label{eqinitialconditions}
\end{eqnarray}
and so forth, reducing the width of the range by factor of five for
each higher order in slow roll. The series is closed to order $M$ by
taking ${}^{M+1}\lambda_{\rm H} = 0$. We use $M = 5$ for the calculations
in this paper. The exact choice of ranges for
the initial parameters does not have a large influence on the result
of the Monte Carlo process, as long as they are chosen such that the
slow roll hierarchy is convergent. As noted above, the form of the
flow equations (\ref{eqfullflowequations}) ensures that the derivative
of ${}^{\ell}\lambda_{\rm H}$ depends only on parameters of order $\ell$
and $\ell + 1$, so this truncation still leads to an {\em exact}
evaluation of the flow equations to infinite order.  We are selecting
a finite subset out of an infinite number of possibilities for initial
conditions, but the background evolution for a given model is
evaluated exactly.\footnote{As usual in all analyses of inflation we
ignore the back reaction of quantum fluctuations on the background
evolution.} 

The result of integrating the flow equations for a particular model is
a ``path'' in the slow roll space parameterized by the
number of e-folds $N$.  Figures 1-3 show examples of paths plotted in
the $\sigma-\epsilon$ plane for different assumptions about the central
values for the parameters $r$, $n$, and $d n / d \log k$. Also plotted is the
corresponding reconstructed potential for each path. Surprisingly, while
quite complex behavior is possible for the slow roll parameters, this
behavior is only weakly reflected in the shape of the potential itself.

\section{Monte Carlo Reconstruction in Practice}

We now describe two concrete applications of this method. First, we
select central values in various regions of the observable parameter
space with the error bars expected from Planck, and reconstruct the
inflationary potential based on this ``synthetic'' data. We find that
Planck will allow us to determine the qualitative form of the
potential, but is insufficiently precise for well-constrained
reconstruction. This is consistent with the conclusions of other
studies \cite{lidsey95,grivell00}.  Second, we choose a region of
parameter space centered on the values for $n$, $r$ and $d n/d \ln{k}$
for a specified potential -- in this case $V(\phi) \propto \phi^4$ --
and apply our reconstruction algorithm with different sized error
bars, thus determining the observational precision that is needed in
order to reliably reconstruct a known potential.

Figure~4 shows the results of two different ``reconstructions'': for
$r=0.02$, $n=0.95$ and $dn/ d \log{k} =0$, and for $r=0$, $n=.93$ and
$d n / d \log k = 0$. In both cases the errors bars are 
those anticipated for the Planck mission's measurement of the spectrum
\cite{kinney98,copeland97}, namely $\delta r \sim 0.01$, $\delta n
\sim 0.01$ and $\delta d n / d \log k \sim  0.01$.  The first case has
a tensor contribution that would be resolved by Planck, and the
normalization of the potential is relatively tightly constrained, $V
\sim 10^{-11} m_{\rm Pl}^4$. The second reconstruction assumes that
the tensor amplitude is not detectable by Planck, with a central value
$r \simeq 0.0$. In this case, the normalization of the potential is
very poorly constrained, although the shape of the potential is
consistent with standard ``small-field'' models
\cite{dodelson97}.  The clear conclusion from these plots, and the
other cases that we have examined, is that Planck may be able to
determine the qualitative features of the potential, but it will not
permit the quantitative reconstruction of the inflationary potential
on its own.

It should not come as any particular surprise that Monte Carlo
reconstruction does not yield a tight constraint on the possible form
of the inflationary potential.  However, the question this immediately 
raises is how much more accurately -- relative to the precision
expected from Planck -- would we need to measure the spectrum in order 
to be able to make a quantitative statement about the functional form
of the potential. We tackle this issue by choosing values of $n$,
$r$ and $d n / d\log{k}$ consistent with a known potential, in this
case the canonical $\lambda \phi^4$ model, which is typical of
``large-field'' models and has a tensor fluctuation amplitude
observable by Planck.  With this assumption, the central values for
the spectral index and tensor-to-scalar ratio are $n=0.943$ and
$r=0.02$ (evaluated 50 e-folding before the end of inflation). For $V
\propto \phi^4$, $d n/d \log{k}$ is close to zero, and we take the
central value to be precisely zero.

\begin{figure}
\centerline{\includegraphics[width=3in]{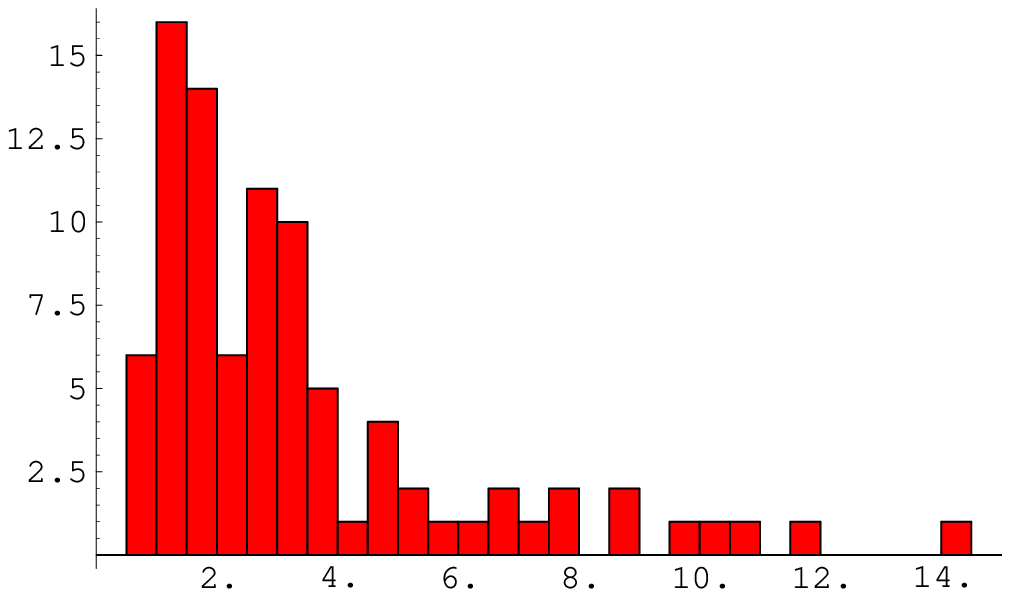}}  
\centerline{\includegraphics[width=3in]{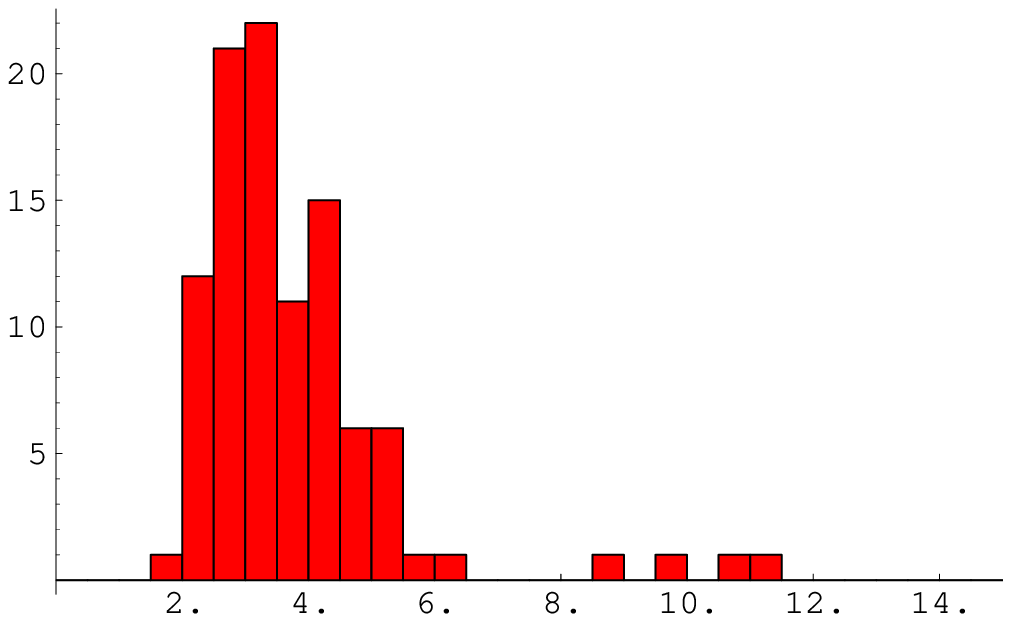}}  
\caption{These two histograms show the values of the power
$m$ (horizontal axes) obtained by fitting Eq.~(\ref{fit}) to 100 potentials
generated by the Monte Carlo reconstruction algorithm, where the measured
spectra are assumed to have the central values predicted by $\lambda
\phi^4$ inflation. The vertical axes indicate the number of models
in a particular bin in $m$. In the top panel we assume that the
error bars on
the spectral parameters are equal to those expected from Planck,
$\delta r \sim 0.01$, $\delta n \sim 0.01$ and $\delta d n / d \log k
\sim 0.01$, whereas the bottom panel corresponds to $\delta r \sim
0.002$, $\delta n \sim 0.002$ and $\delta d n / d \log k \sim 0.005$.
In the former case, the functional form of the potential cannot be
meaningfully recovered, but in the lower case the results are
consistent with $V \propto \phi^4$.  In both plots we have dropped a
few cases for which the least squares solver did not converge.}
\end{figure}

Using the above choices for the central values, we have performed
Monte Carlo reconstruction with Planck-sized errors bars and with
$\delta n = 0.002$, $\delta r = 0.002$ and $\delta d n/d\log{k} =
0.005$. The second choice leads to a box in the $(n,r)$ plane that is
25 times smaller than the bound predicted for Planck and thus
represents a substantial improvement in experimental precision. We
divided the $d n/d \log{k}$ bound by 2 rather than 5 solely for
computational convenience. The second reconstruction involved
computing approximately 90 million ``trial'' inflationary models in
order to filter out 100 solutions that passed through the specified
window in parameter space. This took several days of CPU time on a
fast desktop machine, and would have taken even longer if we had
applied the same scaling to $\delta d n/d\log{k}$ as we did to $\delta
n$ and $\delta r$.

If we plot graphs to analogous to those in Fig.~4 for these two
reconstructions we find that, unsurprisingly, the overall shape of
both sets of reconstructed potentials are consistent, but that there
is less spread in the set with the smaller error bars. However, to
determine whether any sort of quantitative reconstruction is possible,
we adopt the prior that the potentials are proportional to 
$(\phi - \phi_0)^m$, where $\phi_0$ is an unknown offset,
 and then perform a least-squares fit to determine the best-fit value 
of $m$.  In practice, it is sufficient to normalize the height of the
reconstructed potential to precisely unity at $\phi=0$ and then fit it to the
following functional form: 
\begin{equation} \label{fit}
(1- c \phi)^m
\end{equation}
This choice forces the fitted (and rescaled) potential to be unity
when $\phi = 0$. Adding normalization as a third parameter to the 
fit does not have a significant impact on the computed values of $m$.

The results for both simulations are shown in Fig.~5. The error bars
expected from the Planck mission do not allow one to conclude that $V
\propto \phi^4$, but the tighter bounds on $n$ and $r$ do rule out values
of $m$ markedly different from 4.\footnote{In reality, if the computed
value of $m$ is significantly different from 4, what we actually learn
is that the potential cannot be well described by Eq.~(\ref{fit}) --
since setting $m$ to a value significantly different from 4 produces
values of $n$ and $r$ well outside our assumed range.}

We thus tentatively conclude that while Planck cannot measure the
perturbation spectrum accurately enough to put even mild quantitative
constraints on the potential, increasing the accuracy with which both
$n$ and $r$ can be recovered by a factor of 5 would allow one to start
making meaningful reconstructions of the potential.

\section{Discussion}

We have presented a novel method for reconstructing the inflationary
potential, based on the powerful flow equation approach to
inflationary dynamics.
This method, which we dub
{\em Monte Carlo reconstruction\/}, differs from other prescriptions
for reconstruction in that it is stochastic in nature and involves
extremely weak prior assumptions about the form of the potential. A
stochastic approach to reconstruction has distinct advantages compared
with, for example, methods which expand the potential in a Taylor
series and fit the coefficients of the Taylor expansion to observable
parameters. In particular, the method can be applied equally well to
both poor data sets and to high-quality data: parameters such as the
tensor spectral index do not enter the reconstruction in any direct
way, making the method very simple and robust. Also, Monte Carlo
reconstruction naturally produces shapes for the potential outside the
region directly constrained by observation. This follows from our
principal assumption, single field inflation, and the attractor
behavior of the inflationary dynamics which tends to ensure that the
reconstructed potentials overlap outside the region in which we have
direct observational input.

We want to make clear what Monte Carlo reconstruction is {\em not}: it
is a stochastic method, but not a statistical one. We do not have a
metric on the space of initial conditions. Consequently, we cannot use
the ``density'' of models in any particular parameter space to infer
the relative likelihood of one parameter region over another.  Plots
of models in either the space of possible potentials or the space of
exponent $m$ (from $\phi^m$, as in Fig.~5) are properly interpreted as
exclusion plots, indicating which regions are either consistent or
inconsistent with the data.  However, we cannot determine the relative
likelihood of different initial points in slow roll space without an
understanding of inflationary initial conditions. Many other
reconstruction attempts truncate the slow roll expansion at second or
third order, but the approach here can be extended to arbitrary order
in slow roll, and we have checked that our results do not depend on
the specific level at which we truncate slow roll.  However, any slow
roll ansatz effectively rules out potentials with small features where
the higher order derivatives of the potential and $H(\phi)$ are large
\cite{wang97}. If the feature is located in the region of the
potential that corresponds to the primordial spectrum, strict limits
can be placed on the size and slope of the feature
\cite{adams01}.   However, if the feature is outside this region then we
will obviously not be able to reconstruct it.

The most straightforward application of Monte Carlo reconstruction is
to simply generate an ensemble of potentials consistent with some
observational constraint. This simple and robust procedure can be
applied to data sets (for example the forthcoming data from the MAP
satellite) for which direct functional reconstruction of the potential
will likely be impossible. While this method suffers from the same
fundamental limitations as any other method, it will allow us to at
least answer qualitative questions about the form of the inflationary
potential: for example, is the potential convex or concave? Higher
quality data will allow for more quantitative constraints on the
potential. Detection of a nonzero tensor component will give
information about the normalization of the potential, to within an
order of magnitude or two with the accuracy projected for Planck. Even
absent a detection of tensor modes, it may be possible to reach
conclusions about the shape of the potential, if not its
normalization. More quantitative information can be gleaned from more
accurate data sets.  For instance, an improvement in parameter
resolution by a factor of five or so over Planck will make information
about the exponent $\phi^m$ of the potential available in at least a
rough sense. Other possible applications would be plotting the results
of the Monte Carlo reconstruction in the space of derivatives of the
potential, $V, V', V''$, and so on, for comparison with
Refs. \cite{hansen01,caprini02}.

Monte Carlo reconstruction is easily generalizable. For example, one
need not use the slow roll approximation to calculate the power
spectrum associated with a particular choice of ``slow roll''
parameters generated by the flow equations. (Note that the slow roll
expansion is completely distinct from the slow roll {\em
approximation}. The solutions we generate for the background evolution
are exact.) For calculating the fluctuation spectrum associated with a
particular path in the parameter space, one could equally well apply
the method of uniform approximations introduced by Habib, et
al. \cite{habib02}. It is also straightforward to solve for the
perturbation spectrum by numerically solving the exact equation for
quantum modes in the inflationary spacetime: as in the calculation of
the potential itself, all the information required to do so is
contained in the solution to the flow equations. Thus the method is
not only exact in principle, but can be made so in practice if
necessary. In addition, the same techniques could be applied to a set
of flow equations based around an expansion other than the Hubble slow
roll expansion of Liddle et al., such as the expansion used in Ref.
\cite{schwarz01}. Doing so would be useful to investigate the
dependence of the reconstructed potentials on the details of the
truncation scheme for the flow equations. We expect this dependence to
be small.

\section*{Acknowledgments}

RE and WHK are supported by ISCAP and the Columbia University Academic
Quality Fund.  ISCAP gratefully acknowledges the generous support of
the Ohrstrom Foundation.  We would like to thank Ed Copeland for
useful discussions. RE thanks the Aspen Center for Physics, where part
of this work was conducted. Some of the computational time used for
these calculations were kindly provided by the High Energy Theory
group at Brown University.

\end{document}